\documentclass{article} 
\usepackage{iclr2022_conference,times}
\usepackage{amsmath,amsfonts,bm}

\def\eqref#1{equation~\ref{#1}}

\def\1{\bm{1}}

\DeclareMathAlphabet{\mathsfit}{\encodingdefault}{\sfdefault}{m}{sl}
\SetMathAlphabet{\mathsfit}{bold}{\encodingdefault}{\sfdefault}{bx}{n}

\usepackage{hyperref}
\usepackage{url}
\usepackage{amsmath,amssymb,amsfonts}
\usepackage{algorithmic}
\usepackage{graphicx}
\usepackage{textcomp}
\hypersetup{colorlinks,linkcolor={blue},citecolor={blue},urlcolor={red}} 
\usepackage{xcolor}
\usepackage{subcaption}
\usepackage{multirow}
\usepackage{varwidth}
\usepackage{tikz}
\def\BibTeX{{\rm B\kern-.05em{\sc i\kern-.025em b}\kern-.08em
T\kern-.1667em\lower.7ex\hbox{E}\kern-.125emX}}

\title{Astronomical Image colorization and up-scaling with generative adversarial networks}

\author{Shreyas Kalvankar, Hrushikesh Pandit, Pranav Parwate, Atharva Patil \& Snehal Kamalapur \\
Department of Computer Engineering\\
K. K. Wagh Institute of Engineering Education and Research\\
Nashik, Maharashtra, India \\
\texttt{\{shreyas.kalvankar, hrushikesh.pandit\}@kkwagh.edu.in} \\
\texttt{\{pranav.parwate, atharva.patil, smkamalapur\}@kkwagh.edu.in} \\
}

\iclrfinalcopy 
\begin{document}
\usetikzlibrary{chains, calc,fit}

\maketitle

\begin{abstract}
Automatic colorization of images without human intervention has been a subject of interest in the machine learning community for a brief period of time. Assigning color to an image is a highly ill-posed problem because of its innate nature of possessing very high degrees of freedom; given an image, there is often no single color-combination that is correct. Besides colorization, another problem in reconstruction of images is Single Image Super Resolution, which aims at transforming low resolution images to a higher resolution. This research aims to provide an automated approach for the problem by focusing on a very specific domain of images, namely astronomical images, and process them using Generative Adversarial Networks (GANs). We explore the usage of various models in two different color spaces, RGB and L*a*b. We use transferred learning owing to a small data set, using pre-trained ResNet-18 as a backbone, i.e. encoder for the U-net and fine-tune it further. The model produces visually appealing images which hallucinate high resolution, colorized data in these results which does not exist in the original image. We present our results by evaluating the GANs quantitatively using distance metrics such as L1 distance and L2 distance in each of the color spaces across all channels to provide a comparative analysis. We use Fréchet inception distance (FID) to compare the distribution of the generated images with the distribution of the real image to assess the model's performance.
\end{abstract}

\section{Introduction}

Colorization of gray scale images with an automated algorithm has been under much research within the machine learning and computer vision communities. Beyond being fascinating from an aesthetic and artificial intelligence perspective, such capability has broad practical applications. It is an area of research that has untapped potential in applications: from black and white photo reconstruction, image augmentation, image enhancement to video restoration for improved interpretability. \\
\hspace*{0.167 in}Image downscaling is an innately lossy process. The principal objective of super resolution imaging is to reconstruct a low resolution image into a high resolution one based on a set of low-resolution images to rectify the limitations that existed while the procurement of the original low-resolution images, insuring better visualization and recognition for scientific and non-scientific purposes. A particularly good up-scaling algorithm will always have some data loss when producing high frequency image due to a downscale-upscale function. Ultimately, even the best up-scaling algorithms are unable to effectively reconstruct non-existing data. Conventional methods rely on low-information and a smooth interpolation between known pixels. These methods can effectively be treated as a convolution with a kernel which encodes no information about the original image. Generative Adversarial Networks (GANs) can be used to hallucinate high frequency data in a super scaled image that does not exist in the smaller image. Even after increasing the resolution, they fail to achieve the desired clarity in the task. By using the above mentioned method, not a perfect reconstruction can be obtained albeit instead a rather plausible guess can be made at what the lost data might be, constrained to reality by a loss function which penalizes deviations from the ground truth image.\\
    \begin{figure}[!htb]
    	\centering
    	\begin{subfigure}[b]{0.45\textwidth}
    		\centering
    		\includegraphics[width=\textwidth]{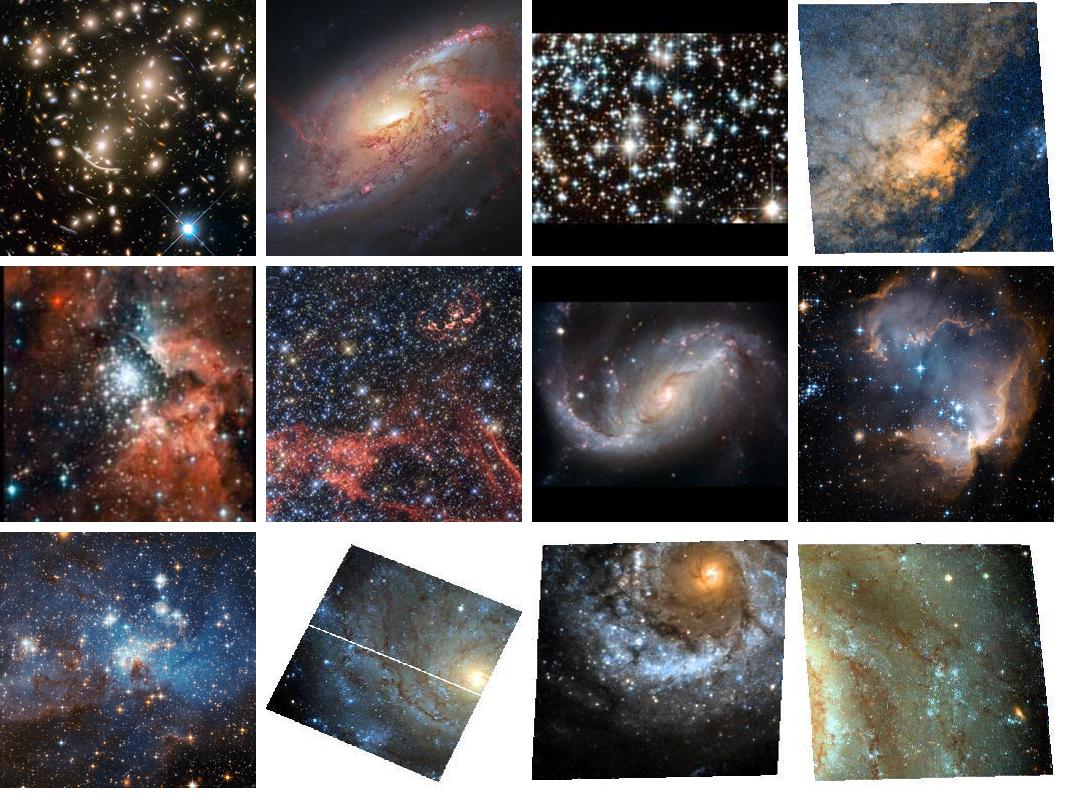}
    		\caption{Ground Truth}
    		\label{fig: main_color_samples}
    	\end{subfigure}
    		\hspace{0.1 in}
    	\begin{subfigure}[b]{0.45\textwidth}
    		\centering
    		\includegraphics[width=\textwidth]{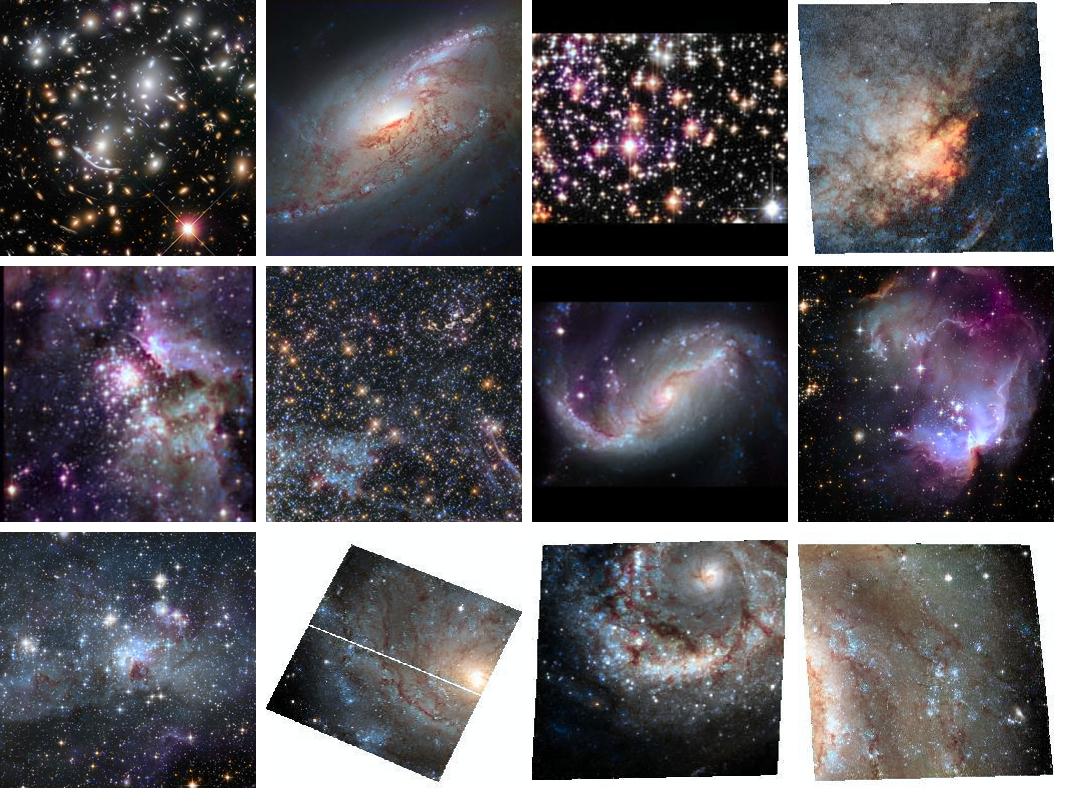}
    		\caption{ResNet18 Full trained L*a*b}
    		\label{fig: resnet18_full}
    	\end{subfigure}
    	\caption{Results of fine-tuned ResNet18 in L*a*b colorspace compared with ground truth images}
    \end{figure}
\hspace*{0.167 in}As noted in \cite{Gao2019astronomical}, a huge number of raw images are unprocessed and unnoticed in the Hubble Legacy Archives. These raw images, typically black and white, low-resolution, are unfit to be shared with the world. It takes huge amounts of hours to process them. This processing is necessary because it's difficult for astronomers to distinguish objects from the raw images and is further made necessary owing to noise from other bodies in the universe, changing optical characteristics in the system and random \& synthetic noise from the sensors in the telescope. Furthermore, for the process of highlighting small features that ordinarily wouldn't be able to be picked out against noise of the image, we need colorization. The processing of the images is so time consuming that the images are rarely seen by human eyes. Not only is new data being continuously produced by Hubble Telescope, but new telescopes are soon to come online. This goes to show that the problem will likely get worse. A simplification of image processing by using artificial image colorization and super-resolution can be done in an automated fashion to make it easier for astronomers to visually identify and analyze objects in Hubble dataset.
    \begin{figure}[!htb]
    	\centering
    	\begin{subfigure}[b]{0.3\textwidth}
    		\centering
    		\includegraphics[width=\textwidth]{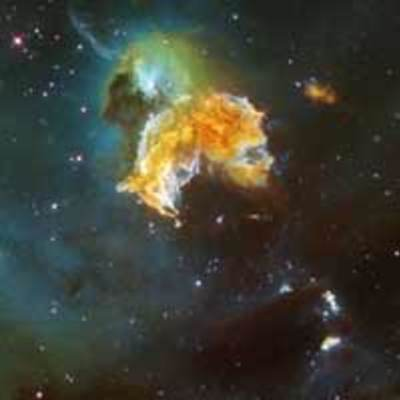}
    		\caption{Original image}
    		\label{fig: original_sample}
    	\end{subfigure}
    	\begin{subfigure}[b]{0.3\textwidth}
    		\centering
    		\includegraphics[width=\textwidth]{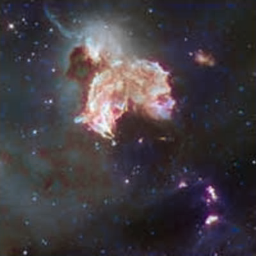}
    		\caption{GAN output}
    		\label{fig: gan_output_sample}
    	\end{subfigure}
    	\caption{Fig \ref{fig: original_sample} is the original image and fig \ref{fig: gan_output_sample} is the image produced by our GAN}
    	\label{fig: sample_comp}
    \end{figure}

\section{Literature Review}
The following section presents a brief survey of existing research and methods explored to approach the problem.
\subsection{Image Colorization}
\subsubsection{Hint Based Colorization}
\hspace*{0.167 in}\citet{levin2004colorization} proposed using colorization hints from the user in a quadratic cost function which imposed that neighboring pixels in space-time with similar intensities should have similar colours. This was a simple but effective method but only had hints which were provided in form of imprecise colored scribbles on the grayscale input image. But with no additional information about the image, the method was able to efficiently generate high quality colorizations. \cite{huang2005edge} addressed the color bleeding issue faced in this approach and solved it using adaptive edge detection. \cite{yatziv2006chrominance} used luminescence based weighting for hints to boost efficiency. \cite{qu2006manga} extended the original cost function to apply color continuity over similar textures along with intensities.

\subsubsection{Deep Colorization}
\hspace*{0.167 in}Owing to recent advances, the Convolutional Neural Networks are a de facto standard for solving image classification problems and their popularity continues to rise with continual improvements. CNNs are peculiar in their ability to learn and differentiate colors, patterns and shapes within an image and their ability to associate them with different classes.\\
    \hspace*{0.167 in}\cite{cheng2016deep} proposed a per pixel training for neural networks using DAISY \citep{tola2008descriptor}, and semantic \citep{long2015semantic} features to predict the chrominance value for each pixel, that used bilateral filtering to smooth out accidental image artifacts. With a large enough dataset, this method proved to be superior to the example based techniques even with a simple Euclidean loss function against the ground truth values.\\
\hspace*{0.167 in}Finally, \cite{dahl2016automatic} successfully implemented a system to automatically colorize black \& white images using several ImageNet-trained layers from VGG-16 \citep{simonyan2015deep} and integrating them with auto-encoders that contained residual connections. These residual connections merged the outputs produced by the encoding VGG16 layers and the decoding portion of the network in the later stages. \cite{he2015deep} showed that deeper neural networks can be trained by reformulating layers to learn residual function with reference to layer inputs. Using this \textit{Residual Connections}, \cite{he2015deep} created the \textit{ResNets} that went as deep as 152 layers and won the 2015 ImageNet Challenge.\\
    
\subsubsection{Generative Adversarial Networks}
\hspace*{0.167 in}\cite{goodfellow2014generative} introduced the adversarial framework that provides an approach to training a neural network which uses the generative distribution of $p_g(x)$ over the input data $x$.\\
    \hspace*{0.167 in}Since it's inception in 2015, many extended works of GAN have been proposed over years including DCGAN \citep{radford2016unsupervised}, Conditional-GAN \citep{mirza2014conditional}, iGAN \citep{zhu2018generative}, Pix2Pix \citep{isola2018imagetoimage}.\\
    \hspace*{0.167 in}\cite{radford2016unsupervised} applied the adversarial framework for training convolutional neural networks as generative models for images, demonstrating the viability of \textit{deep convolutional generative adversarial networks}.\\
    \hspace*{0.167 in}DCGAN is the standard architecture to generate images from random noise. Instead of generating images from random noise, Conditional-GAN \citep{mirza2014conditional} uses a condition to generate output image. For e.g. a grayscale image is the condition for colorization of image. Pix2Pix \citep{isola2018imagetoimage} is a Conditional-GAN with images as the conditions. The network can learn a mapping from input image to output image and also learn a separate loss function to train this mapping. Pix2Pix is considered to be the state of the art architecture for image-image translation problems like colorization.
    
\subsection{Image Upscaling}
\subsubsection{Frequency-domain-based SR image approach}
\hspace*{0.167 in} \cite{tsai1984multiframe} proposed the frequency domain SR method, where SR computation was considered for the noise free low resolution images. They transformed the low resolution images into Discrete Fourier transform (DFT) and further combined it as per the relationship between the aliased DFT coefficient of the observed low resolution image and that of unknown high resolution image. Then the output is transformed back into the spatial domain where a higher resolution is now achieved.\\
    \hspace*{0.167 in} While Frequency-domain-based SR extrapolates high frequeny information from the low resolution images and is thus useful, however they fall short in real world applications.\\
\subsubsection{The interpolation based SR image approach}
\hspace*{0.167 in} The interpolation-based SR approach constructs a high resolution image by casting all the low resolution images to the reference image and then combining all the information available from every image available.
    The method consists of the following three stages
    (i) the registration stage for aligning the low-resolution input images,
    (ii) the interpolation stage for producing a higher-resolution image, and
    (iii) the deblurring stage which enhances the
    reconstructed high-resolution image produced in the step (ii).
    
    However, as each low resolution image adds a few new details before finally deblurring them, this method cannot be used if only a single reference image is available.\\
    
\subsubsection{Regularization-based SR image approach}
\hspace*{0.167 in} Most known Bayesian-based SR approaches are maximum likelihood (ML) estimation approach  and maximum a posterior (MAP) estimation approach.\\
    \hspace*{0.167 in}  While \cite{Brian1996ML} proposed the first ML estimation based SR approach with the aim to find the ML estimation of high resolution image, some proposed a MAP estimation approach. MAP SR tries to takes into consideration the prior image model to reflect the expectation of the unknown high resolution image.\\
    
\subsubsection{Super Resolution - Generative Adversarial Networks (SR-GAN)}
\hspace*{0.167 in} The Genrative Adversarial Network \citep{goodfellow2014generative}, has two neural networks, the Generator and the Discriminator. These networks compete with each other in a zero-sum game.
    \cite{ledig2017photorealistic} introduced SRGAN in 2017, which used a SRResNet to upscale images with an upscaling factor of 4x. SRGAN is currently the state of the art on public benchmark datasets.
    
\section{Methodology}
\begin{figure}[!htb]
        \centering
        \resizebox{\textwidth}{!}{
        \input{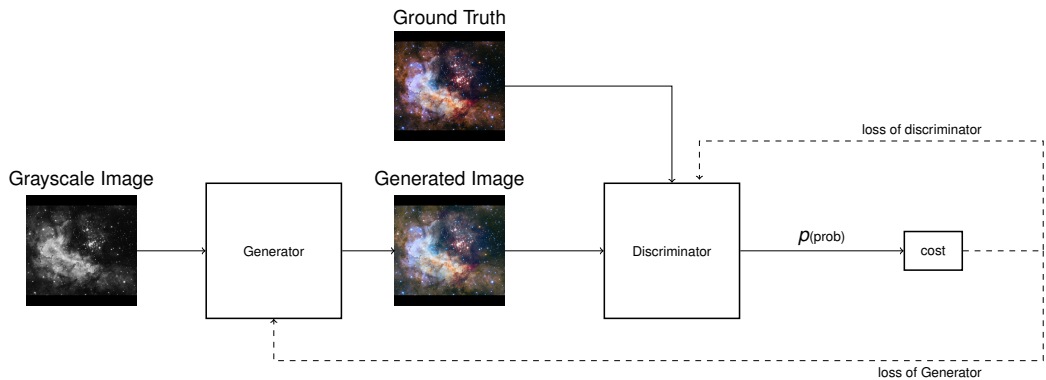}}
        \caption{Approach for image colorization using conditional GANs. A sample grayscale image is provided as an input to the generator and the generated image along with the ground truth is fed to the discriminator. The probability \textit{p} is the probability that the discriminator thinks the two images are similar.}
        \label{fig:gan_visualization}
    \end{figure}
\subsection{Data collection}
    \hspace*{0.167 in}We started by scraping data off the Hubble Legacy Archive\footnote{Based on observations made with the NASA/ESA Hubble Space Telescope, and obtained from the Hubble Legacy Archive, which is a collaboration between the Space Telescope Science Institute (STScI/NASA), the Space Telescope European Coordinating Facility (ST-ECF/ESA) and the Canadian Astronomy Data Centre (CADC/NRC/CSA).}. The scrapper tool\footnote{Fork: https://github.com/obi-wan-shinobi/hla-scraper} which was used, courtesy of \cite{Gao2019astronomical}\footnote{Original repository: https://github.com/KMarshland/hla-scraper}, scraped off hundreds of thousands of colorized images the archive has available. The Hubble Legacy archive is slow and produces grainy images with lots of noise and majority are unprocessed. A filter for M101 (Messier 101) galaxy rendered more than 80 thousand images with a 1 degree difference between consecutive right ascension. The data acquired was large and there was no particularly efficient way to clean it without human investment. Cleaning tens of thousands of images by handpicking noiseless and well colored images is time consuming. For training the SRGAN, high resolution and well colored images were needed.\\
    \hspace*{0.167 in}Consequently, we scraped the Hubble Heritage project instead. The Hubble Heritage project releases the highest-quality astronomical images. They are all stitched together, colorized and processed to eliminate noise. Hubble Heritage then selects the best, most striking of these for public release. However, there were only $\sim150$ of these images that are actually useful. We scraped images from the main Hubble website as well so as to increase the amount of data we had. This provided another $\sim1000$ images approx. Limited by our computational resources, it was decided that the images will be in the dimensions of $256\times 256$ pixels and contains 3 channels namely Red, Green, Blue (RGB).\\
    
    \subsection{Image Color Space}
\hspace{0.167 in}An RGB image is essentially a rank 3 tensor of height, width and color. The data is represented in RGB color space which has 3 numbers for every pixel indicating the amount of \textit{Red, Green, and Blue} values the pixel has. In \textbf{L*a*b} color space, we have three numbers for each pixel but these numbers have different meanings.L, the first channel, has the \textbf{Lightness} of each pixel encoded and when we visualize this channel it appears as a black and white image. 
    \begin{figure}[!htb]
    \centering
    \begin{subfigure}[b]{0.9\textwidth}
        \centering
    	\includegraphics[width=\textwidth]{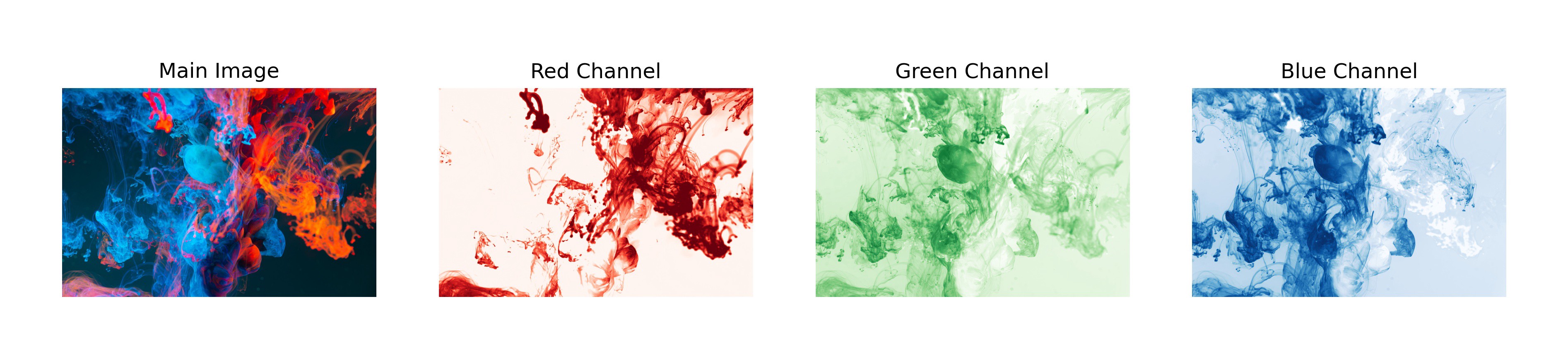}
    	\caption{Reg, Green and Blue channels of an image (Main image by \href{https://unsplash.com/@aznbokchoy}{Lucas Benjamin})}
    	\label{rgb_colorspace}
    \end{subfigure}
    \begin{subfigure}[b]{0.9\textwidth}
        \centering
    	\includegraphics[width=\textwidth]{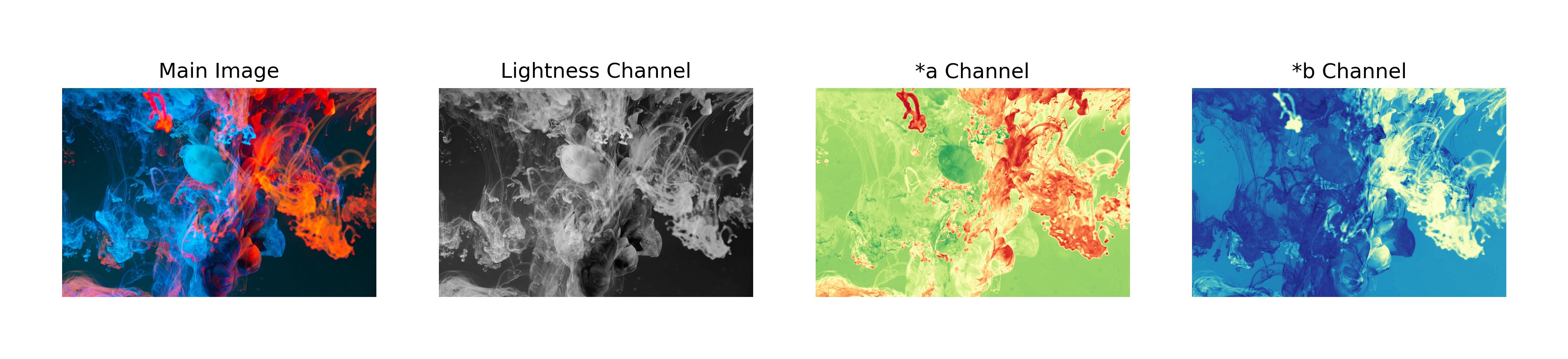}
    	\caption{Lightness, *a and *b channels of the L*a*b colorspace}
    	\label{lab_colorspace}
    \end{subfigure}
    \caption{Color spaces}
    \end{figure}
    The \textbf{*a and *b} channels encode in them exactly how much \textbf{green-red} and \textbf{yellow-blue} each pixel is, respectively.\\
\hspace*{0.167 in}\cite{guadarrama2017pixcolor,isola2018imagetoimage} proposed the use of L*a*b color space instead of RGB for the colorization task. Intuitively, to the train a model for colorization, a grayscale image should be fed to it and we hope that it colors it. In the L*a*b colorspace, the model is fed the L channel, which is essentially a grayscale image, and we perform computations to predict the other two channels (*a and *b). After the predictions, we concatenate the channels and get a colorful image. In case of RGB, there is a need to explicitly convert the three channels down to 1 to make it a grayscale image and then feed it to the network hoping that it predicts three numbers per pixel. This is an unstable task due to sheer increase in volume of combinations from two numbers to three. We train models using both color spaces and compare their performance.
    \subsection{Mathematical Model}
    Generative modeling uses Generative Adversarial Networks (GANs) for hallucinating an output vector space. A mathematical description is provided for a GAN below.
    \subsubsection{Generative Adversarial Networks}
\hspace*{0.167 in}A generative network, $G$, is supposed to learn the underlying distribution of a latent space, Y. Instead of visually assessing the quality of network outputs and judge how we can adapt the network to produce convincing results, we incorporate automatic tweaking during training by introducing a discriminative network $D$. The network D takes in both the fabricated outputs generated by G and real inputs from the underlying distribution Y. The network produces a probability of the image belonging to the real or fabricated space.\\
\hspace*{0.167 in}Let $x \in X$ be a low resolution or a grayscale image and $y \in Y$ be it's underlying distribution from the latent space Y. Generator $G$ takes in input $x$ and produces an output $\hat{y}$. We define the mapping $x \rightarrow \hat{y}$ in the following manner:
      $$G(x) = \hat{y}$$
    
\hspace*{0.167 in}The discriminative network D is fed the fabricated mapping $x \rightarrow \hat{y}$ and the underlying distribution of $x$ i.e. $y \in Y$. The network D then produces a result that is a probability distribution of the input space indicating the class of the image that it thinks the input belongs to. We define this as:
     $$D\big(G(x),y\big) = p$$
     where $p \in (0,1)$ is the probability that the image is fabricated or real.
\hspace*{0.167 in}With conditional GAN, both generator and discriminator are conditioning on the input $x$. Let the generator be parameterized by $\theta_g$ and the discriminator be parameterized by $\theta_d$. The minimax objective function can be defined as:
    	\begingroup\makeatletter\def\f@size{7}\check@mathfonts \[
    		\min_{\theta_g}\max_{\theta_d}\Big[\mathbb{E}_{x,y\sim p_{data}} \log D_{\theta_d}(x,y) + E_{x\sim p_{data}} \log(1 - D_{\theta_d}(x, G_{\theta_g}(x))\Big]
    	\] \endgroup
        \normalsize
    Where, $G_{\theta_{g}}$ is the output of the generator and $D_{\theta_d}$ is the output of the discriminator.
    We're currently not introducing any noise in our generator to keep things simple for the time being. Also, we consider $L1$ difference between input $x$ and output $y$ in generator. On each iteration, the discriminator would maximize $\theta_d$ according to the above expression and generator would minimize $\theta_g$ in the following way:
    	\[
    		\min_{\theta_g}\Big[-\log(D_{\theta_d}(x,G_{\theta_g}(x)))+\lambda \Vert G_{\theta_g}(x) - y \Vert_1 \Big]
    	\]
    
    \subsection{Transferred learning and model tweaking}
    \hspace*{0.167 in}\cite{isola2018imagetoimage} proposed a general solution to many image-to-image translation tasks. We propose a similar methodology as proposed in the paper with a few minor tweaks, so as to significantly reduce the amount of training data needed and minimize the training time to achieve similar results. Initially, we propose the use of a U-net for the generator. A pre-trained ResNet-18 network was used as the \textit{encoder} for the U-net with half of it's layers clipped off so as to get feature abstractions of the intermediate layers. \cite{ledig2017photorealistic} showed that training the generator separately in a supervised, deterministic manner helps it generalize the mapping from the input space to outputs. The idea solves the problem of \textit{"The blind leading the blind"} that is persistent in most image-to-image translation tasks to date. \\
    \hspace*{0.167 in}We pre-train our generator independently using the imagenet weights over the Common Objects in Context (COCO) dataset. Unlike adversarial training, this phase was supervised and the loss function used was \textit{mean absolute error} or the \textit{L1 Norm} from the target image. Though trained deterministically,the problem of rectifying incorrect predictions still persists due to constraints over the convergence of loss metric. To combat this, this trained generator was trained once again in an adversarial fashion to help generalize it further. We hypothesize that re-training in an adversarial fashion will further rectify the subtle color differences that \textit{mae} couldn't solve.\\
    \hspace*{0.167 in}A pre-trained ResNet-50 with imagenet weights was used as the discriminator with last few layers clipped off. The discriminative network used is something called a "Patch Discriminator" proposed by \cite{isola2018imagetoimage}. In a \textit{vanilla} discriminator proposed by \cite{goodfellow2014generative}, the network produces a scalar output, representing the probability that the data $x$ belongs to the input distribution and not the generator distribution $p_g$. \cite{isola2018imagetoimage} proposed a modification to the discriminative network so as to produce, instead of a single scalar, a vector of probabilities representing different localities of the input distribution (image) $x$. For instance, a discriminator producing a $50 \times 50$ vector represents the probabilities every receptive field that is covered by the output vector. Thus, we can localize the corrections in the image that the generator should make.\\
    \hspace*{0.167 in}Finally, the trained generator and trained discriminator are fine-tune to fit on our data which is rather small in size compared to the previous datasets. These networks are trained in an adversarial fashion using the conditional GAN objective function \citep{isola2018imagetoimage} with some noise introduced as the L1 norm of generated image tensor to the target image tensor. The reason behind this is to train the generator using an adversarial component while trying to minimize the Manhattan distance between the generator output and target vector space.
    \subsection{Experimental Setup}
    \hspace*{0.167 in}We use a mini-batch gradient descent with a batch size of 10, 16 and 32 for different iteration with Adam that has $\beta_1 = 0.5$ and $\beta_2 = 0.999$ as momentum. The generator and discriminator have a learning rate of $2e-4$ which remains constant throughout the training. We train the model with different epochs ranging between 20,50 and 100, saving the best model weights determined by L1 norm between the output of the generator and the target image. We use early stopping with a patience value of 10. The image size is $256\times256$. The implementation uses Python, numpy, Tensorflow and tf-keras. It takes about 24 hours to complete training over the COCO dataset and about 12 to 13 hours to fit the model on given astronomical data on a NVIDIA Tesla K80 GPU.
    
\section{Results and Discussions}
    \hspace*{0.167 in}In the following section, we compare the results of the implemented architectures and evaluate their performance. The evaluation is done qualitatively as well as quantitatively by comparing the performance of each model using L1 and L2 distance between the predictions and targets. Though unreliable, this method provides us with a somewhat decent ground to perform a comparative study and evaluate the reliability of such metrics on GAN evaluation compared to qualitative, visual evaluation. The performance of the GANs is also measured by using Fréchet inception distance (FID) as an attempt to quantify how close the produced images are to the original ones.\\
    \hspace*{0.167 in}To evaluate the model performance by virtue of convergence of the objective function, we plot the generator loss throughout the training along with the discriminator.\\
    \begin{figure}[!htb]
    	\centering
    	\begin{subfigure}[b]{0.35\textwidth}
    		\centering
    		\includegraphics[width=\textwidth]{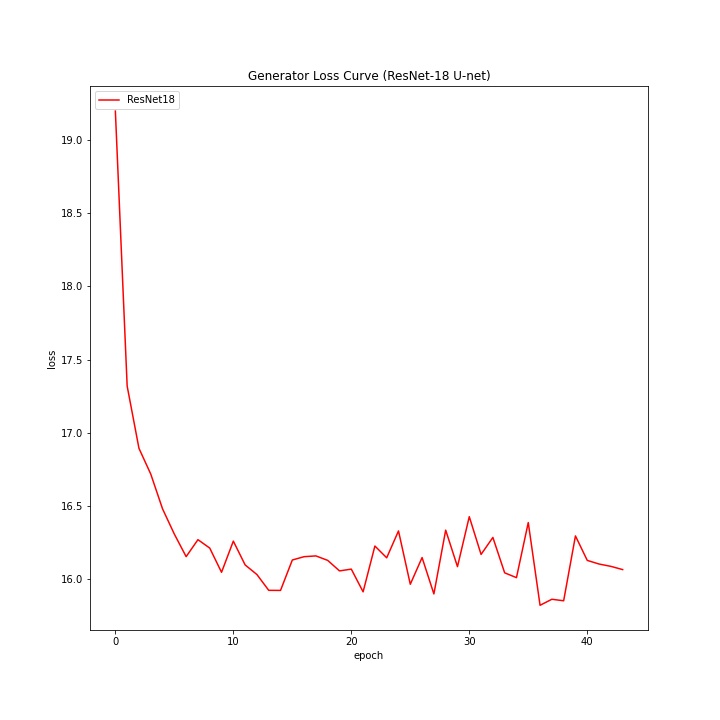}
    		\caption{ResNet18 U-net Generator}
    		\label{fig: resnet_loss}
    	\end{subfigure}
    		\hspace{0.1 in}
    	\begin{subfigure}[b]{0.35\textwidth}
    		\centering
    		\includegraphics[width=\textwidth]{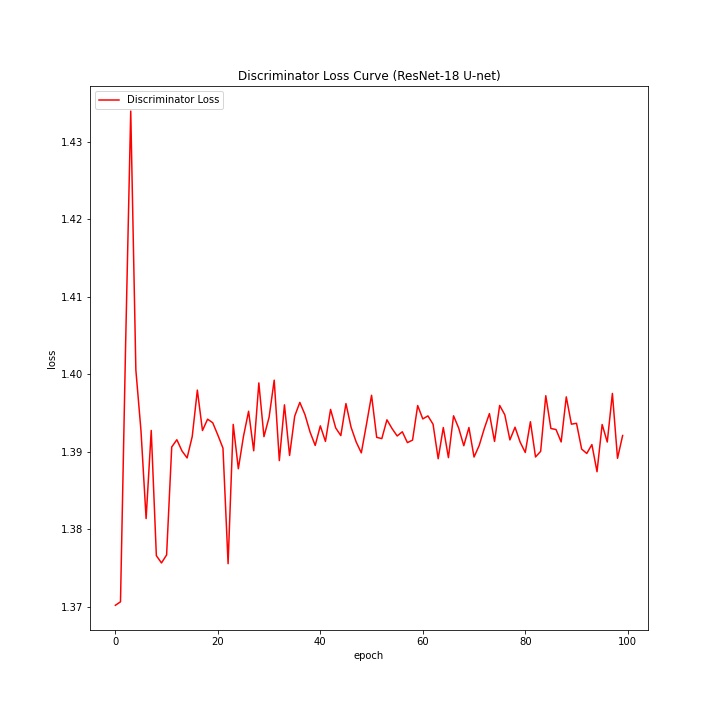}
    		\caption{ResNet50 discriminator loss}
    		\label{fig: resnet50_disc_loss}
    	\end{subfigure}
    	\caption{Curve plots for ResNet18 U-net \& ResNet50 discriminator}
    	\label{fig: resnet18_curves}
    \end{figure}
    Figure \ref{fig: resnet_loss} and \ref{fig: resnet50_disc_loss} shows that the generator converges with a little instability. The discriminator on the other hand oscillates because of the convergence that the generator shows. GAN losses are pretty non-intuitive but we draw a few conclusions from the graphs. This pattern of oscillation and convergence repeats when networks are trained in an adversarial fashion. It is observed that the generator with a pre-trained ResNet-18 for its backbone converges decently in the beginning but later shows spikes when nearing the end of training loop. The convergence doesn't signify whether the model is predicting expected results. The loss just converges to the minimum value that the cost function descends to, permitted by the \textit{lr}. It would normally mean that the GAN has found some optimum point in the vector space that is at the lowest potential and can't decrease any further, meaning the GAN has learned enough. Due to the sheer no. of dimensions, owing to the high amount of trainable variables, such combinations, where the function converges, can be high in volume. Thus, these numbers don't provide any better understanding of the bias or variance the model is facing. We also discover that even if the loss hasn't converged well it doesn't necessarily mean that the model hasn't learned anything. On visual inspection, the generated results show similar distribution to the ground truth, even with high generator losses. This might be due to presence of a content loss parameter in the adversarial loss function which is also minimizing the L1 norm between the predictions and target.\\
\hspace*{0.167 in}The discriminator shows an increase in the objective loss in the initial epochs, settles down in the later phase around a point, converging to some permanent number and oscillating around it. We assume this point to be a point of stability between the two networks as the networks are in a constant adversarial battle, meaning if one performs better, the other is bound to perform worse.
    \subsection{Image Colorization}
\begin{figure}[!htb]
    	\centering
    	\begin{subfigure}[b]{0.35\textwidth}
    		\centering
    		\includegraphics[width=\textwidth]{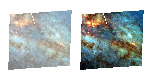}
    		\caption{Sample 2}
    		\label{fig: overfit1}
    	\end{subfigure}
    		\hspace{0.1 in}
    	\begin{subfigure}[b]{0.35\textwidth}
    		\centering
    		\includegraphics[width=\textwidth]{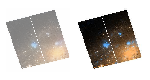}
    		\caption{Sample 2}
    		\label{fig: overfit2}
    	\end{subfigure}
    	\caption{Performance of Basic U-net over samples different from the dataset. The left part is the prediction, right part is the ground truth}
    	\label{fig: u-net_overfit}
    \end{figure}

\begin{figure}[!htb]
    	\centering
    	\begin{subfigure}[t]{0.3\textwidth}
    		\centering
    		\includegraphics[width=\textwidth]{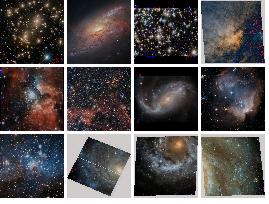}
    		\caption{Basic U-net}
    		\label{fig: u-net_samples}
    	\end{subfigure}
    	\begin{subfigure}[t]{0.3\textwidth}
    		\centering
    		\includegraphics[width=\textwidth]{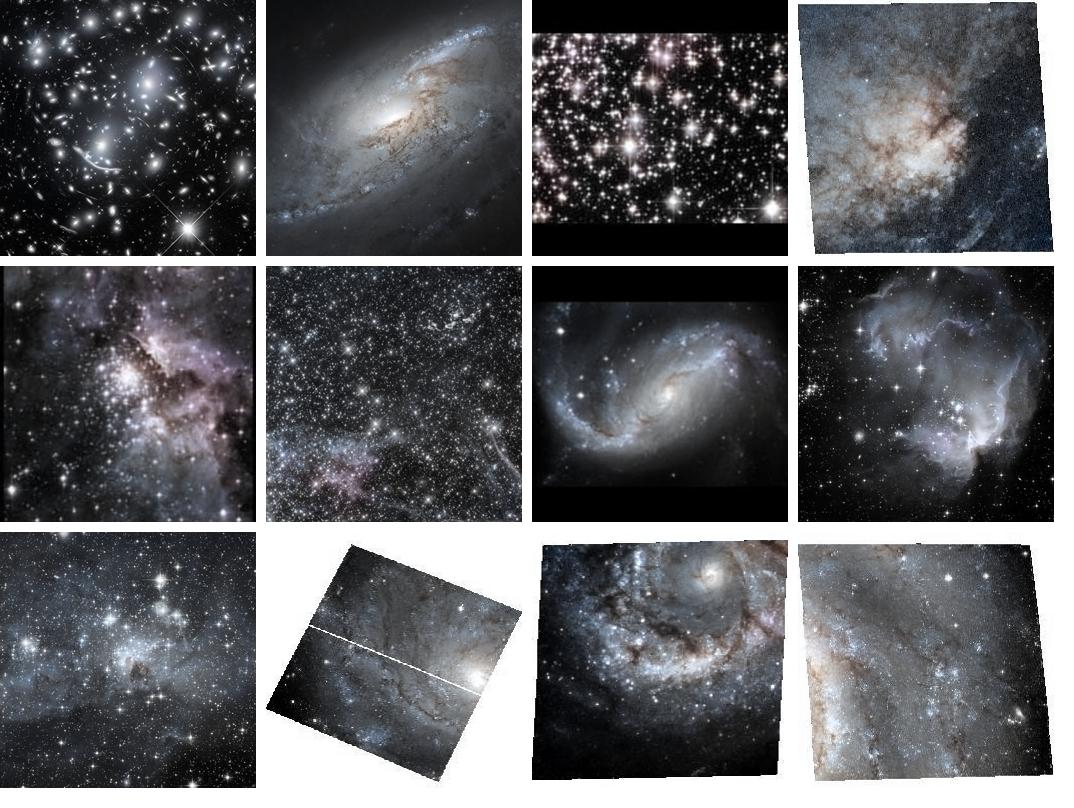}
    		\caption{ResNet18 pre-trained L*a*b}
    		\label{fig: resnet18_pre}
    	\end{subfigure}
    	\begin{subfigure}[t]{0.3\textwidth}
    		\centering
    		\includegraphics[width=\textwidth]{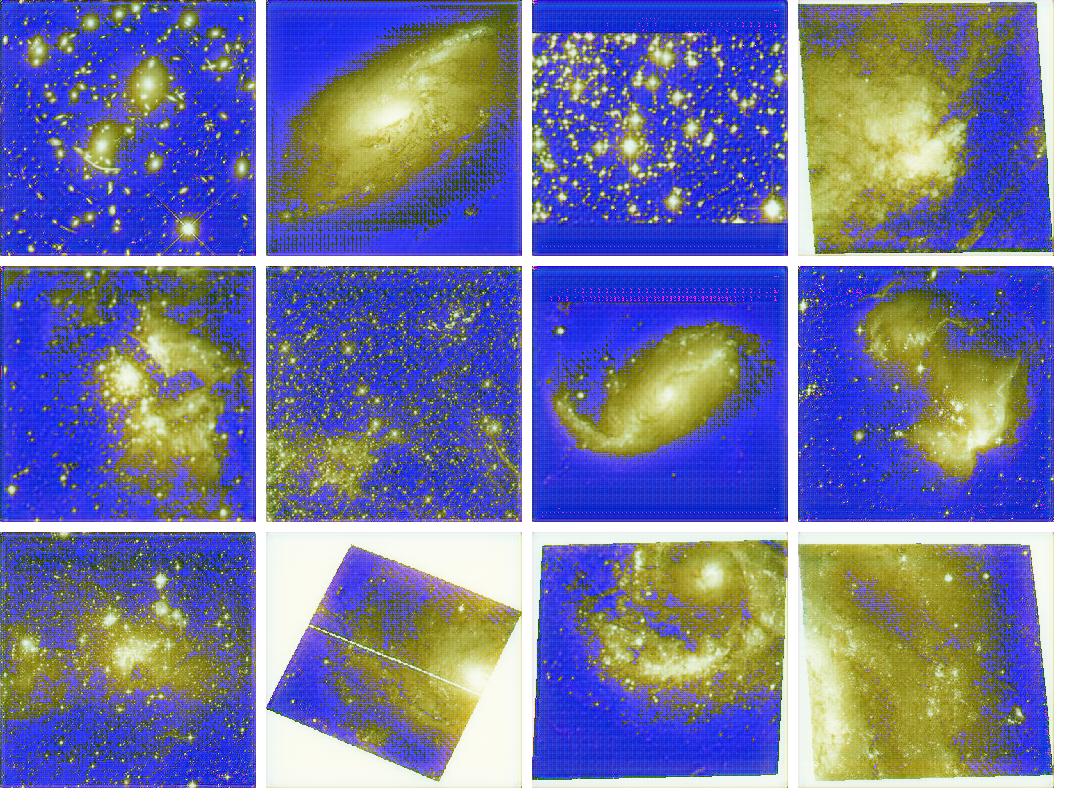}
    		\caption{ResNet18 pre-trained RGB}
    		\label{fig: resnet18_rgb}
    	\end{subfigure}
    	\caption{Results of colorization models (from top to bottom): Basic U-net, ResNet18 U-net pre-trained in L*a*b color space, ResNet18 U-net pre-trained in RGB color space}
    	\label{fig: comparisons}
    \end{figure}

    \hspace*{0.167 in}We present a comparative study of the following models: Basic U-net generator with a residual VGG16 discriminator, hereafter referred to as Basic U-net. A modified U-net with pre-trained ResNet18 as it's backbone. We evaluate the model by training it in RGB colorspace to predict 3 \textit{nos.} for every pixel and in L*a*b colorspace where the model predicts the a and b channel alone. We then further fine-tune this model to the task of colorizing astronomical images.
    \begin{figure}[!htb]
    \centering
    	\includegraphics[width=0.5\textwidth]{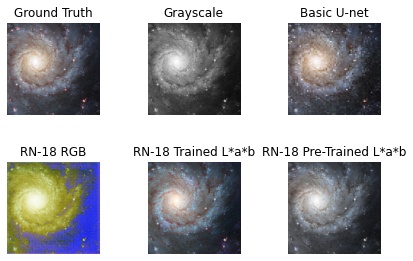}
    	\caption{Results of the colorization study.From top left, (1) the ground truth and (2) input image. The next images belong to the output of following, serially: Basic U-net, ResNet18 in RGB, ResNet18 fine-tuned in L*a*b, ResNet18 pre-trained U-net in L*a*b}
    	\label{color_results}
    \end{figure}
    Figure \ref{color_results} shows that on a particular example, the Basic U-net performs better at predicting results that closely map to the ground truth even if the fine-tuned ResNet-18 shows better results for larger set of inputs. It can also be observed that the Basic U-net architecture does a decent job of faking the sharpness in the original image and that makes the image appear more realistic as compared to rather blurry outputs from the other networks. \\
   \hspace*{0.167 in}We also observe that the model trained in the RGB color space performs poorly at predicting values of all color channels and that results in channel bias (blue in this case). This causes the output, though reconstructed quite accurately, to have a varied color pattern with high emphasis on one color channel. This might be caused by the deep layers which result in diminishing gradients for certain colors over time, causing the model to be strongly biased towards the blue color in this case. An increase in the volume of training data and some random pixel shuffling in forward propagation might solve this problem. \\
    \hspace*{0.167 in}The pre-trained ResNet18 U-net performs decently with the weights gathered by training it over the COCO dataset. The model still lacks the specific coloring intuition in astronomical images and plainly colors specific parts of the images in light colors, leaving majority of the image unaltered. This causes the images to retain grayness.  \\  
    \hspace*{0.167 in}Figure \ref{fig: comparisons} shows how the other models perform in different color spaces. We observe that the Basic U-net model performs good at predicting the outputs but suffers at predicting the brightness level of pixels. The model seems to be overfitting on the dataset and suffers a high variance on output as demonstrated in figure \ref{fig: u-net_overfit}.
    
    Figure \ref{fig: main_color_samples} shows the ground truth images and figure \ref{fig: resnet18_full} shows the predictions of the ResNet18 full trained model in the L*a*b color space. This model seems to perform best, visually. 
    
    To quantitatively estimate the model performance, we measure the L1 and L2 distance between the predictions and the ground truth images in both RGB as well as L*a*b colorspace. 
    \begin{table}[!htb]
\centering
\setlength{\tabcolsep}{4pt} 
\renewcommand{\arraystretch}{1.5} 
\footnotesize
\begin{tabular}{c | c | c | c}
        \hline
        \textbf{Model} & \textbf{Color Space} & \textbf{
        L1 Distance} & \textbf{L2 Distance}\\
        \hline
        ResNet-18 (Pre-trained) & L*a*b & 64.5409 & 2.77\\
        ResNet-18 (Fine-tuned) & L*a*b & 65.1119 & 2.62\\
        ResNet-18 (Pre-trained) & RGB & 125.5554 & 9.04\\
        Vanilla/Basic U-net & RGB & 77.3273 & 3.986\\
        \hline
    \end{tabular}
\caption{Colorization: Average per-pixel L1 \& L2 distance between generated images and ground truth}
\label{tab:colorization_results}
\end{table}
    Figure \ref{color_results} shows that the RGB color space performs poorly on the task of colorization. In terms of the L1 distance, the best performance is achieved on the ResNet18 U-net with pre-trained weights. This goes to prove the unreliability of distance metrics in model performance evaluation. The model, although quantitatively, performs better than the fine-tuned model but in reality, fails to produce images that might be visually appealing. The L2 distance metric shows how the fine-tuned model might, in reality, be fitting slightly better over the data to predict correct color combination as its output. We still consider the results from the fine tuned model to be better than the other performing models and further investigate the per-channel predictions by the model.
    \begin{table}[!htb]
\centering
\setlength{\tabcolsep}{4pt} 
\renewcommand{\arraystretch}{1.5} 
    \begin{tabular}{c | c | c | c}
        \hline
        \textbf{Distance} & \textbf{Red} & \textbf{
        Green} & \textbf{Blue}\\
        \hline
        L1 Norm & 64.6078 & 38.5994 & 92.1283\\
        L2 Norm & 3.4531 & 1.0253 & 3.8046\\
        \hline
    \end{tabular}
\caption{Channel wise averages of ResNet-18 finetuned network}
\label{tab:channel_wise_results}
\end{table}
    Table \ref{tab:channel_wise_results} shows the RGB channel averages of the outputs produced by the fine tuned model. It can be observed that the feature embeddings produced by the model in *a, *b color spaces maps to the green channel with the least error. This might be indicative that the model is performing poorly on images that have a high content of red-blue colors. 
    \begin{table}[!htb]
\centering
\setlength{\tabcolsep}{4pt} 
\renewcommand{\arraystretch}{1.5} 
    \begin{tabular}{c | c c | c c}
        \hline
        \hline
        \multirow{2}{*}{Model} & 
            \multicolumn{2}{c|}{L1 Distance} & \multicolumn{2}{c}{L2 Distance}\\
            & a & b & a & b\\
        \hline
        ResNet-18 (Pre-trained)  & 0.00588 &  0.00501 & 0.01565 & 0.03208\\
        ResNet-18 (Fine-tuned) & 0.00257 & 0.00727 & 0.01721 & 0.03316\\
        \hline
    \end{tabular}
\caption{L1 \& L2 channel wise distances in L*a*b colorspace}
\label{tab:lab_color_results}
\end{table} 
    
In table \ref{tab:lab_color_results}, L1 distance, we can see the performance of fine-tuned model to be better in channel *a but poor in *b compared to the pre-trained model. The L2 distance metric rules out the possibility of the fine-tuned model performing better than the pre-trained model. In reality, the fine-tuned model is orders of magnitude better at predicting output abstractions with the L channel as its input, thus contradicting the quantitative results.
    \begin{table}[!htb]
\setlength{\tabcolsep}{4pt} 
\renewcommand{\arraystretch}{1.5} 
\small
\centering
    \begin{tabular}{c | c c c| c c c}
        \hline
        \hline
        \setlength{\tabcolsep}{2pt}
        \multirow{2}{*}{Model} & 
            \multicolumn{3}{c|}{\footnotesize{L1 Distance}} & \multicolumn{3}{c}{\small L2 Distance}\\
            & \footnotesize{R} & \footnotesize{G} & \footnotesize{B} & \footnotesize{R} & \footnotesize{G} & \footnotesize{B}\\
        \hline
        \footnotesize{ResNet-18 (PT)}  & \footnotesize{69.2401} & \footnotesize{36.5852} & \footnotesize{87.7973} & \footnotesize{3.2890} & \footnotesize{1.0439} & \footnotesize{3.9785}\\
        \footnotesize{ResNet-18 (FT)} & \footnotesize{64.6078} & \footnotesize{38.5994} & \footnotesize{92.1283} & \footnotesize{3.4531} & \footnotesize{1.0253} & \footnotesize{3.8046}\\
        \hline
    \end{tabular}
\caption{L1 \& L2 channel wise distances in RGB colorspace (PT refers to Pre-trained and FT refers to Fine-tuned)}
\label{tab:rgb_color_results_pre_fine}
\end{table}
    We can observe the comparative results of ResNet18 pre-trained model and the fine-tuned model. It can be concluded from the observations that the fine-tuned model performs better at predicting the Red color channel but suffers in Green and Blue.
    \begin{table}[!htb]
\centering
\setlength{\tabcolsep}{4pt} 
\renewcommand{\arraystretch}{1.5} 
\begin{tabular}{c | c }
        \hline
        \textbf{Model} & \textbf{FID}\\
        \hline
        ResNet-18 (Pre-trained) & 66.3568\\
        ResNet-18 (Fine-tuned) & 42.1593\\
        U-net & 152.7216\\
        \hline
    \end{tabular}
\caption{Colorization: Fréchet inception distance between model outputs and original images}
\label{tab:colorization_FID}
\end{table}
    We use Fréchet inception distance (FID) to evaluate the distribution of generated images with the distribution of real images. Table \ref{tab:colorization_FID} helps us see that the fine-tuned U-net with ResNet-18 as it's backbone achieves the least FID score. This shows that while this model is very adept at hallucinating images it's still not able to predict accurate color values. 
    \subsection{Image Super Resolution}
    \hspace*{0.167 in}We implement the basic SR-GAN proposed by Ledig and train it to improve super-resolution task. We compare the trained model with pre-trained SRGAN model, EDSR-GAN proposed by \cite{lim2017enhanced} and WDSR-GAN proposed by \cite{yu2018wide}.
    \begin{figure}[!htb]
    \centering
    	\includegraphics[width=0.5\textwidth]{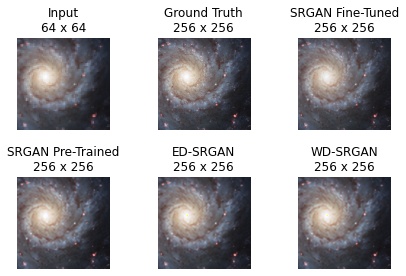}
    	\caption{Results of the super-resolving models. The first two images in the top row are the input image and ground truth respectively. The next images belong to the output of following, serially: Fine-tuned SRGAN, pre-trained SRGAN, EDSR, WDSR}
    	\label{fig: sr_results}
    \end{figure}
    Figure \ref{fig: sr_results} shows how the networks perform on the task of predicting pixels while upscaling of the image. All the networks perform really well and its difficult to distinguish the outputs visually. It may be observed that the fine-tuned network produces images that look slightly better than the other counter-parts. The best performing network seems to be the WDSR-GAN with a very little error in predicting the output. 
    \begin{figure}[!htb]
    	\centering
    	\begin{subfigure}[b]{0.4\textwidth}
    		\centering
    		\includegraphics[width=\textwidth]{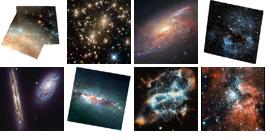}
    		\caption{Input Distribution}
    		\label{fig: sr_input_samples}
    	\end{subfigure}
    	\begin{subfigure}[b]{0.4\textwidth}
    		\centering
    		\includegraphics[width=\textwidth]{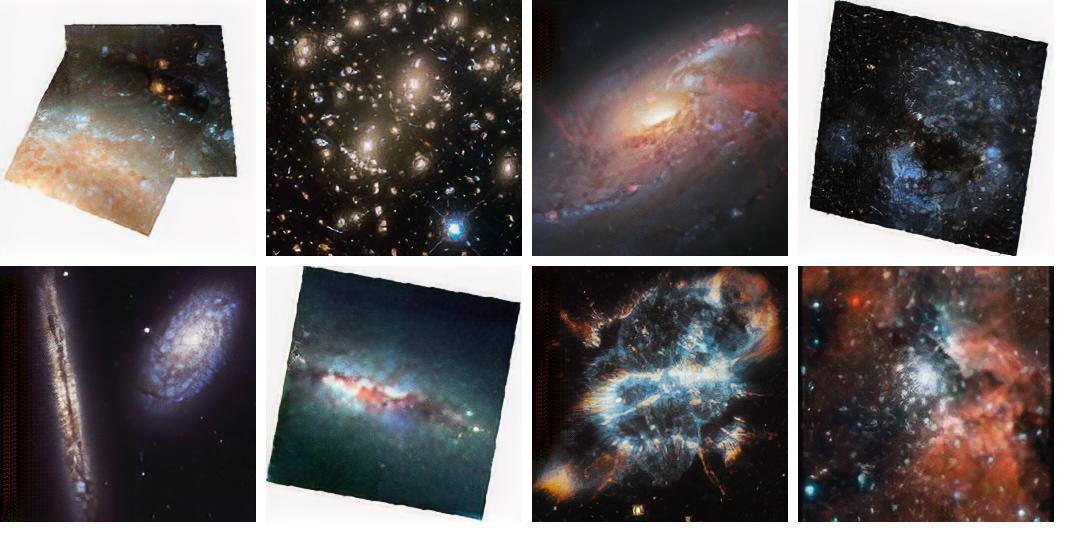}
    		\caption{SRGAN fine-tuned output}
    		\label{fig: sr__outputs}
    	\end{subfigure}
    	\caption{Results of SRGAN. \ref{fig: sr_input_samples} shows the input data and \ref{fig: sr__outputs} shows the corresponding output}
    	\label{fig: sr_comparisons}
    \end{figure}
    It is evident that the model produces acceptable results on visual inspection. The main reason behind this might, again, be the random pixel shuffling between every upscaling pass. 
    As opposed to colorization, super-resolution needs a quantitative estimation to determine which model performs best among the give models. 
    \begin{table}[!htb]
\centering
\setlength{\tabcolsep}{10pt} 
\renewcommand{\arraystretch}{1.5} 
    \begin{tabular}{c | c | c }
        \hline
        \textbf{Model} & \textbf{
        L1 Distance} & \textbf{L2 Distance}\\
        \hline
        Ledig SRGAN (Fine-tuned) & 87.1090 & 3.754\\
        Ledig SRGAN (Pre-trained) & 114.8043 & 5.953\\
        ED-SRGAN & 80.9414 & 3.684\\
        WD-SRGAN & 79.7262 & 3.627\\
        \hline
    \end{tabular}
\caption{Super-Resolution: Per-pixel Average L1 \& L2 distance between generated images and ground truth}
\label{tab:Super-resolution_results}
\end{table}
    Table \ref{tab:Super-resolution_results} shows the L1 and L2 distances of predicted results by each model with the ground truth image. It is observed that Ledig's SRGAN, after a bit of fine-tuning performs really well in comparison to the the pre-trained version. To further improve this, \cite{lim2017enhanced} proposed an optimized version of SRGAN by removing the unnecessary modules in the conventional resnets and showed that Enhanced Deep Residual Networks performed better at upscaling task. When trained in an adversarial manner, the ED-SRGAN performs better than the traditional SRGAN. \cite{yu2018wide} further improved the idea by increasing the widening factor ($\times 2$ and $\times 4$) to ($\times 6$ and $\times 9$). This Wide Activation Deep Super Resolution network further improved the performance for single image super-resolution. When we implement this in an adversarial manner, we achieve excellent results. It is evident from the results that the best performing network is the WDSR. 
    \begin{table}[!htb]
\centering
\setlength{\tabcolsep}{4pt} 
\renewcommand{\arraystretch}{1.5} 
\begin{tabular}{c | c }
        \hline
        \textbf{Model} & \textbf{FID}\\
        \hline
        Ledig SRGAN (Fine-tuned) & 75.5873\\
        Ledig SRGAN (Pre-trained) & 89.8443\\
        ED-SRGAN & 159.5770\\
        WD-SRGAN & 127.2920\\
        \hline
    \end{tabular}
\caption{Super-resolution: Fréchet inception distance between model outputs and original images}
\label{tab:super-resolution_FID}
\end{table}
    
    Having said that, if we look at the FID score listed in table \ref{tab:super-resolution_FID}, we notice that the lowest FID score is given by the fine-tuned SRGAN (which is ideally what we expect), but this just proves our previous claim that GANs are particularly good at hallucinating things rather than predicting them. 
    \section{Summary and Conclusion}
    \hspace*{0.167 in}The project mainly tackles the problem of colorizing astronomical images and super-resolving them for astronomical inspection. We explore various methodologies proposed till date that efficient colorize and super-scale images with results that are significantly closer to the ground truth distribution. We scrape the data from Hubble Legacy Archive, Hubble Main website and Hubble Heritage project and created a filtered and clean dataset of 4700 images. We split the data and use 500 images, roughly 10\% of the data for testing purposes. To compensate for the lack of data, we implement several pre-trained architectures and fine-tune their abstractions over our dataset to find the most effective solution.\\
    \hspace*{0.167 in}For the colorizing model, we explore usage of U-net architectures starting from a basic U-net model and experiment with different color spaces and empirically confirm the superiority of L*a*b color space in image colorization problem. We use the state of the art ResNet-18 to provide as a backbone of the encoder and build a U-net around it. The pre-trained network over COCO dataset in RGB colorspace produces significantly weaker results as compared to the subsequent network in L*a*b colorspace. The best performing model turns out to be the ResNet18 U-net which is fine-tuned over our particular dataset to produce appealing and similar results to the ground truth.\\
    \hspace*{0.167 in}The Super-resolution model is based largely on the SRGAN proposed by \cite{ledig2017photorealistic}. We use the generator weights and sample results from the training set to inspect the results. It is found that the model performs really well on the pre-trained weights and we decide to fine-tune it to our application. After fine-tuning the model, we train other state of the art single image super-resolution models such as EDSR \citep{lim2017enhanced} and WDSR \citep{yu2018wide}. These provide further insights into the problem and simultaneously, improve the results.\\
    \hspace*{0.167 in}While studying and improving the performance of these models, we explore performance metrics of GANs and evaluation methodologies implemented to test out conditional GANs. It is evident that the loss curves of generator and discriminator do not provide us with any intuition about the model performance. We also discover that standard distance metrics cannot be used to evaluate GANs and quantitative methods that exist to evaluate GANs are unreliable. We prove so by contradiction of qualitative samples and quantitative measurements of the best performing architecture for colorization models. However, we observe that quantitative estimation is quite reliable for the problem of single image super-resolution and can be helped to determine which model is better suited for the task.
    
    \section{Future Scope}
    \hspace*{0.167 in}Though we obtain moderately good results, a vast amount of algorithms still remain unscratched. A more powerful model such as SE-ResNext, EfficientNet and more such state-of-the-art models can be implemented and trained over millions of images from the Imagenet. With even more hardware resources and availability of data, we can explore computationally heavy models for a better approximation. With help of an image stitching algorithm, produced images can be stitched together to generate large scale astronomical images for scientific study. Colorization can be improved by the virtue of exploring different loss functions using weighted losses to reduce loss problem for low saturation regions. We can introduce a gradient penalty for the SRGAN architecture and include the WGAN \citep{arjovsky2017wasserstein} which will stabilize the discriminator by enforcing conditions which result in a Lipschitz constant $<1$, so that it will stay within the space of 1-Lipschitz function. Progressively growing GANs \citep{karras2018progressive} can be applied so that the dimensions can be further improved with more stability and greater sharpness.

\bibliography{Astronimical_Image_Colorization_and_upscaling}
\bibliographystyle{iclr2022_conference}

\end{document}